\documentclass[twocolumn,aps,superscriptaddress,showpacs]{revtex4}
\usepackage{amssymb}
\usepackage{amsmath}
\usepackage{graphicx}
\usepackage[normalem]{ulem}
\usepackage[dvips]{color}
\usepackage{multirow}
\usepackage{appendix}
\usepackage{CJK}
\usepackage{multirow}
\usepackage{booktabs}
\usepackage{threeparttable}
\makeatletter

\newcommand{\Rmnum}[1]{\expandafter\@slowromancap\romannumeral #1@}

\makeatother

\makeatletter % `@' now normal "letter"
\@addtoreset{equation}{section}
\makeatother  % `@' is restored as "non-letter"

\begin{document}

\title{Effects of hadronic mean-field potentials on Hanbury-Brown-Twiss correlations in relativistic heavy-ion collisions}
\author{Chun-Jian Zhang}
\affiliation{Shanghai Institute of Applied Physics, Chinese Academy
of Sciences, Shanghai 201800, China}
\affiliation{University of Chinese Academy of Sciences, Beijing 100049, China}
\author{Jun Xu\footnote{corresponding author: xujun@sinap.ac.cn}}
\affiliation{Shanghai Institute of Applied Physics, Chinese Academy
of Sciences, Shanghai 201800, China}
\date{\today}

\begin{abstract}
With the parameters fitted by the particle multiplicity, the energy density at chemical freeze-out, and the charged particle elliptic flow, we have studied the effects of the hadronic mean-field potentials on the Hanbury-Brown and Twiss (HBT) correlation in relativistic heavy-ion collisions based on a multiphase transport model. The hadronic mean-field potentials are found to delay the emission time of the system and lead to large HBT radii extracted from the correlation function. Effects on the energy dependence of $R_o^2-R_s^2$ and $R_{o}/ R_{s}$ as well as the eccentricity of the emission source are discussed. The HBT correlations can also be useful in understanding the mean-field potentials of protons, kaons, and antiprotons as well as baryon-antibaryon annihilations.
\end{abstract}

\maketitle

\section{Introduction}
\label{introduction}

The Beam-Energy-Scan (BES) program at the Relativistic Heavy-Ion Collider (RHIC) has been performing lower-energy relativistic heavy-ion collisions in order to map out the phase diagram of quantum chromodynamics (QCD) at lower temperatures and finite baryon chemical potentials~\cite{MMA10}. At the center-of-mass (C.M.) energy from 39 to 7.7 GeV, the hadronic evolution, which was traditionally treated as the final-state interaction (FSI) of relativistic heavy-ion collisions, is expected to become more and more important in the whole dynamics of the collision. Based on the Boltzmann transport framework, such FSI includes the $``$hard$"$ process of elastic and inelastic scatterings among hadrons, and the $``$soft$"$ process of the evolution of hadrons in their mean-field potentials. For instance, the hadronic mean-field potentials can be partially responsible for the splitting of elliptic flow between particles and their antiparticles at RHIC-BES energies~\cite{JX12}. Understanding the hadronic evolution and constraining precisely the hadronic mean-field potentials are important in describing the dynamics in relativistic heavy-ion collisions at RHIC-BES energies and relevant for extracting reliable information of the QCD phase diagram.

The particle interferometry serves as a useful tool in understanding the space-time and momentum correlations encoding the dynamics as well as the interaction among particles. This technique was first proposed by Hanbury-Brown and Twiss (HBT)~\cite{HBT56} in order to measure the angular diameter of bright visual stars from coherent photon beams. Later, this method was widely
applied in many areas of elementary physics, such as the electron correlations in semiconductors and insulators~\cite{MSH86} as well as analyzing the fermionic statistics of electrons~\cite{MS99} and the Bose-Einstein condensation of ultracold atoms~\cite{SM05}.
The HBT analysis method has also been applied in nuclear physics studies especially for heavy-ion collisions (see, e.g., Refs.~\cite{Boa90,WB92,UAW99,UH99,Wei00,MAL05}). The interferometry of hadrons, especially that of pions, is important in understanding the evolution of the quark-gluon plasma (QGP) formed in ultrarelativistic heavy-ion collisions~\cite{SP84,SP86,GB88,SP90a,DH96}.

The mean-field potential is expected to affect the HBT correlation
in two ways. First, the mean-field potential, which is related to
the equation of state (EOS) of the medium, affects the whole
evolution of the system, and thus the size of the emission source as
well as the emission time of particles at kinetic freeze-out. It was
previously illustrated by an ultrarelativistic quantum molecular
dynamics model that the HBT radii may change after introducing
mean-field potentials~\cite{LQF08}. Second, since the HBT
correlation represents the correlation among the space, time, and
momentum at the emission stage, particles affected by different
mean-field potentials are emitted at different times and are
expected to have different correlations. As shown by an
isospin-dependent Boltzmann-Uehling-Uhlenbeck transport model,
correlation functions of neutrons and protons are sensitive to their
different mean-field potentials~\cite{CLW03}.

Based on a multiphase transport (AMPT) model, we have incorporated the mean-field potentials in the hadronic phase~\cite{JX12}. In the present work, we study the effects of hadronic mean-field potentials on the HBT correlation in relativistic heavy-ion collisions at the RHIC-BES energies. We find that both the HBT correlations and HBT radii are largely modified by the mean-field potentials. It is thus suggested to include the hadronic mean-field potential effects in order to extract accurately the properties of the emission source in relativistic heavy-ion collisions in transport model studies, especially at RHIC-BES energies. On the other hand, the HBT correlations for specific hadron species could be used to constrain their mean-field potentials.

The rest of the paper is organized as follows. In Sec.~\ref{models} we briefly describe the model and formalism used in the present
study, i.e., the multi-phase transport model, the hadronic mean-field potentials, and the femtoscopic formalism. The corresponding analyses and results are discussed in detail in Sec.~\ref{results}. A summary is given in Sec.~\ref{summary}.

\section{Model and formalism}
\label{models}

\subsection{AMPT model}
\label{ampt}

In the following, we briefly review the basic structure of the string-melting AMPT model~\cite{Lin05} used in the present study. The initial phase-space information of partons is generated from melting hadrons produced by the Heavy-Ion Jet INteration Generator (HIJING)
model~\cite{Wang91} through the Lund string fragmentation model, with the fragmentation function $f(z) \propto z^{-1}(1-z)^{a}\exp(-bm_{\perp }^{2}/z)$,
where $a$ and $b$ are parameters, and $z$ is the light-cone momentum fraction of the produced hadron of transverse
mass $m_{\perp}$ with respect to that of the fragmenting string. The evolution of the partonic phase is then
modeled by Zhang's Parton Cascade (ZPC) model~\cite{Zhang98}, where the interaction between quarks or antiquarks is effectively
described by two-body elastic scatterings with the total cross section $\sigma \approx \frac{9\pi \alpha _{s}^{2}}{2\mu ^{2}}$, where $\alpha_s$ is the strong coupling constant, and
$\mu$ is the screening mass in the partonic matter. The freeze-out time of each parton is in principle determined by its last scattering, while in the present work we stop the partonic evolution artificially in order to reproduce the energy density near quark-hadron phase transition, or presumedly, that at chemical freeze-out. When the partonic evolution ends, a spatial coalescence is used for hadronization, where a pair of quark and antiquark close in coordinate space can form a meson, three quarks (antiquarks) close in coordinate space can form a baryon (antibaryon), and the hadron species is determined by the flavors of its valence
quarks and their invariant mass. After hadronization, the hadronic evolution is described by a relativistic transport (ART) model~\cite{Li95,Lin05},
where elastic and inelastic scatterings among hadrons including baryon and antibaryon productions and annihilations as well as resonance decays of hadrons are properly treated. We have applied the recent corrections~\cite{Xu16} to the inelastic channels in the ART model and ensure the charge conservation during the whole hadronic evolution. The hadronic mean-field potentials in the ART model are incorporated through the test-particle method~\cite{Won82,JX12}, i.e., the local phase-space distribution is calculated from averaging parallel events with the same impact parameters, and the mean-field potentials for baryons, kaons, and pions as well as their antiparticles will be detailed in the next subsection. A hadron is considered as kinetically frozen-out if the distance between its current position and the expected position from free streaming is less than 0.01 fm, and the femtoscopic analysis is based on the freeze-out phase-space distribution of hadrons. Note that the criterion for hadron freeze-out is the same as the last scattering if there are only scatterings among hadrons. Since the soft mean-field potentials can further affect the hadron momentum after its last scattering, the hadron freeze-out times are expected to be later.

\subsection{Hadronic mean-field potentials}
\label{HMFP}

The mean-field potentials for nucleons and antinucleons are taken as those from the relativistic mean-field model~\cite{GQ94}, i.e.,
\begin{eqnarray}
U_{N,\bar{N}}\left ( \rho_{B},\rho_{\bar{B}} \right )=\Sigma_{s}\left (\rho_{B},\rho_{\bar{B}} \right )\pm \Sigma_{v}^{0}\left ( \rho_{B},\rho_{\bar{B}} \right ),
\end{eqnarray}%
in terms of the nucleon scalar self-energy $\Sigma_{s}\left (\rho_{B},\rho_{\bar{B}} \right )$ and the time component of the vector self-energy
$\Sigma_{v}^{0}\left ( \rho_{B},\rho_{\bar{B}} \right )$ in hadronic matter of baryon density $\rho_{B}$ and
antibaryon density $\rho_{\bar{B}}$. The $``+"$ and $``-"$ signs are for nucleons and antinucleons, respectively. The detailed form of $\Sigma_{s}$ and $\Sigma_{v}^0$ can be found in Ref.~\cite{GQ94}. It should be noted that
nucleons and antinucleons contribute both positively to $\Sigma_{s}$ but positively and negatively to $\Sigma_{v}^0$, respectively,
as a result of the G-parity invariance. Since only the light quarks in baryons and antibaryons contribute to the scalar and vector self-energies
in the mean-field approach, the potentials of strange baryons and antibaryons are reduced relative to those of nucleons and antinucleons
according to the ratios of their light-quark numbers.

The mean-field potentials for kaons and antikaons in the nuclear medium can be obtained based on the chiral effective Lagrangian~\cite{GQ97}
through $U_{K,\bar{K}}=\omega_{K,\bar{K}}-\omega_{0}$, with
\begin{eqnarray}
\omega_{K,\bar{K}}=\sqrt{m_{K}^{2}+p^{2}-a_{K,\bar{K}}\rho_{s}+\left ( b_{K}\rho_{B}^{net} \right )^{2}}\pm b_{K}\rho_{B}^{net}
\end{eqnarray}%
and $\omega_{0}=\sqrt{m_{K}^{2}+p^{2}}$, where $m_{K}$ is the kaon mass and $a_{K} = 0.22$ GeV$^{2}$fm$^{3}$, $a_{\bar{K}} = 0.45$ GeV$^{2}$fm$^{3}$,
and $b_{K} = 0.33$ GeVfm$^{3}$ are empirical parameters taken from Ref.~\cite{GQ97}. In the above, the $``+"$ and $``-"$ signs are for kaons and antikaons, respectively, $\rho_{s}$ is the scalar density determined from $\rho_{B}$ and $\rho_{\bar{B}}$ through the effective Lagrangian used for describing the properties of nuclear matter~\cite{GQ94}, and $\rho_{B}^{net} = \rho_{B} - \rho_{\bar{B}}$ is the net baryon density.

The mean-field potentials for pions are related to their self-energies $\Pi _{s}^{\pm 0}$ according to $U_{\pi^{\pm 0} }=\Pi_{s}^{\pm 0}/\left ( 2m_{\pi} \right )$,
where $m_\pi$ is the pion mass. The contribution of the pion-nucleon $s$-wave interaction to the pion self-energy is taken from Ref.~\cite{NK01} up to the two-loop order in chiral perturbation theory. In isospin asymmetric nuclear matter of proton density $\rho_{p}$ and neutron density $\rho_{n}$,
the resulting self-energies for $\pi^{-}$, $\pi^{+}$, and $\pi^0$ are expressed respectively by
\begin{eqnarray}
%\begin{split}
\Pi_{s}^{-}\left ( \rho_{p},\rho_{n} \right ) &=&
\rho_{n}\left [ T_{\pi N}^{-}- T_{\pi N}^{+} \right ] -\rho_{p}\left [ T_{\pi N}^{-}+ T_{\pi N}^{+} \right ] \notag \\
&+& \Pi_{\rm rel}^{-}\left ( \rho_{p},\rho_{n} \right )+\Pi_{\rm cor}^{-}\left ( \rho_{p},\rho_{n} \right ), \\
\Pi_{s}^{+}\left ( \rho_{p},\rho_{n} \right ) &=& \Pi_{s}^{-}\left ( \rho_{n},\rho_{p} \right ), \\
\Pi_s^0(\rho_p,\rho_n)&=&-(\rho_p+\rho_n)T^+_{\pi N}+\Pi^0_{\rm cor}(\rho_p,\rho_n).
%\end{split}
\end{eqnarray}%
In the above, $T^{\pm}$ are the isospin-even and isospin-odd pion-nucleon $s$-wave scattering matrices, $\Pi_{\rm rel}^{-}$ is due to the relativistic correction, and $\Pi_{\rm cor}^{-}$ and $\Pi_{\rm cor}^{0}$ are the contribution from the two-loop order in chiral perturbation theory. Their detailed expressions can be found in Ref.~\cite{NK01}.
For nucleon resonances and strange baryons in hadronic matter, we simply extend the above result by treating them as neutron- or proton-like according to their isospin structure and light-quark numbers. The contributions of antiprotons and
antineutrons in hadronic matter are similar to those of neutrons and protons, respectively, as a result of the G-parity invariance. Only the pion-nucleon $s$-wave mean-field potential is incorporated in the present study, while the studies of pion-nucleon $p$-wave interaction and the effects in heavy-ion collisions can be found in Refs.~\cite{GE75,JX81,LX93,Zha17}.

The mean-field potential in the baryon- and neutron-rich hadronic matter formed in relativistic heavy-ion collisions at RHIC-BES energies is strongly attractive for antibaryons, weakly attractive for baryons, strongly attractive for antikaons, weakly repulsive for kaons, slightly attractive for $\pi^{+}$ and $\pi^0$, and slightly repulsive for $\pi^{-}$. In the dominating low-density phase, the potentials for baryons and pions are mostly attractive. For quantitative values of the mean-field potentials for nucleons, kaons, and pions as well as their antiparticles, we refer the reader to Fig.~1 of Ref.~\cite{JX12}.

\subsection{Femtoscopic analyses}
\label{FF}

In order to obtain the HBT correlation, we use the CRAB
(Correlation After Burner) model~\cite{CRAB} and the Lednick$\acute{y}$ and Lyuboshitz analytical model~\cite{LL82} to analyze the phase-space distribution of hadrons at their kinetic freeze-out obtained by AMPT. Both models include FSI which serves as the $``$afterburner$"$ effect after kinetic freeze-out, and the formalisms to obtain the correlation function are detailed in~\ref{LLAM}. In the present study, the CRAB model is used to evaluate the direction-averaged and three-dimensional correlation functions for pions, in order to extract the bulk properties of the system at kinetic freeze-out, while the Lednick$\acute{y}$ and Lyuboshitz analytical model is used to analyse the HBT correlations for protons, kaons, and their antiparticles, in order to compare their different correlation functions due to different mean-field potentials.

%\subsection{Bertsch-Pratt parametrization and Coulomb interactions}
%\label{BP}

The direction-averaged properties of the emission source can be extracted by fitting the HBT correlation function as
\begin{equation}\label{1d}
C(q_{\rm inv}) = (1-\lambda) + \lambda K_{\rm coul}(q_{\rm inv}) \left ( 1+e^{ -q_{\rm inv}^{2}R_{\rm inv}^{2}} \right ),
\end{equation}
where $q_{\rm inv}$ is the magnitude of the relative momentum, $\lambda$ is the parameter characterizing the degree of chaotic or coherent emission~\cite{JA05}, and the function
$K_{\rm coul}(q_{\rm inv})$ represents the Coulomb correction~\cite{YMS98,MGB91}. In the three-dimensional analysis, the relative momentum $\vec{q}$ of particle pairs is decomposed according to the Bertsch-Pratt $``$out-side-long$"$ (o-s-l) convention~\cite{YMS98}, i.e., $q_{l}$ along the beam direction, $q_{o}$ parallel to the transverse momentum of the pair
$\vec{k}_{T}=\left ( \vec{p}_{1T} + \vec{p}_{2T}\right )/2$ with $\vec{p}_{1T}$ and $\vec{p}_{2T}$ being the transverse momenta of the two particles, and $q_{s}$ perpendicular to $q_{l}$ and $q_{o}$. The relative momentum is expressed in the longitudinal co-moving system in which the longitudinal component of the pair velocity vanishes.  The three-dimensional properties of the emission source can be extracted by fitting the correlation function as
\begin{eqnarray}\label{3d}
\begin{footnotesize}
\begin{split}
C(\vec{q}) =& (1-\lambda) + \lambda K_{\rm coul}(q_{\rm inv}) \\
& \times  \left ( 1+e^{ -q_{o}^{2}R_{o}^{2}-q_{s}^{2}R_{s}^{2}-q_{l}^{2}R_{l}^{2}-2q_{o}q_{s}R_{os}^{2}-2q_{o}q_{l}R_{ol}^{2}} \right ).
\end{split}
\end{footnotesize}
\end{eqnarray}%
The $R_{ol}^{2}$ term vanishes in the analysis for midrapidity particles, while the $R_{os}^{2}$ term vanishes in the azimuthal-integrated analysis. In azimuthal-differential analysis, their dependence on the azimuthal angle $\Phi$ for a given $k_{T}$ can be expanded as
\begin{eqnarray}\label{az}
\begin{footnotesize}
\begin{split}
&R_{\mu}^{2}\left ( k_{T}, \Phi \right ) \\&
= R_{\mu,0}^{2} \left ( k_{T} \right )+ 2\sum_{n=2,4,6...}R_{\mu,n}^{2} \left ( k_{T} \right )\cos \left ( n\Phi \right ) \quad
\left  (\mu =o, s, l, ol \right ), \\
&R_{\mu}^{2}\left ( k_{T}, \Phi \right ) \\&
=2\sum_{n=2,4,6...}R_{\mu,n}^{2} \left ( k_{T} \right )\sin \left ( n\Phi \right ) \quad \left   (\mu =os\right).
\end{split}
\end{footnotesize}
\end{eqnarray}%
In the above, $R_{\mu,n}^{2}$ is the $n$th-order Fourier coefficients, which can be obtained by averaging the $\Phi$ dependence as
\begin{eqnarray}
R_{\mu,n}^{2}\left ( k_{T} \right )=\left\{
\begin{aligned}
& \left \langle R_{\mu}^{2}\left ( k_{T},\Phi \right)  \cos(n\Phi) \right \rangle,   \quad  \left   ( \mu =o, s, l \right ) \\
& \left \langle R_{\mu}^{2}\left ( k_{T},\Phi \right)  \sin(n\Phi) \right \rangle.   \quad  \left    ( \mu =os \right)
\end{aligned}
\right.
\end{eqnarray}%
The zeroth-order Fourier coefficient are expected to be nearly
identical to the radii extracted from an azimuthal-integrated
anallysis. In the present analysis, $\Phi$ angle is calculated from
$\Phi =\phi_{pair} -\psi_{2}$, where $\phi_{pair}$ is the azimuthal
angle of the average pair transverse momentum vector $\vec{k}_{T}$,
and $\psi_{2}$ is the azimuthal angle of the second-order
event plane.

\section{Results and discussions}
\label{results}

\subsection{Parameter Settings}
\label{para}

In the present study, we investigate the hadronic mean-field potential effects on the HBT correlation based on the hadron phase-space distribution at their kinetic freeze-out generated by the string-melting AMPT model. In order to have a reliable description of the dynamics in $^{197}$Au + $^{197}$Au collisions at RHIC-BES energies, we fit the parameters to reproduce the particle multiplicity, the energy density at chemical freeze-out, and the elliptic flow.

\begin{figure}[htbp]
\centerline{\includegraphics[scale=0.3]{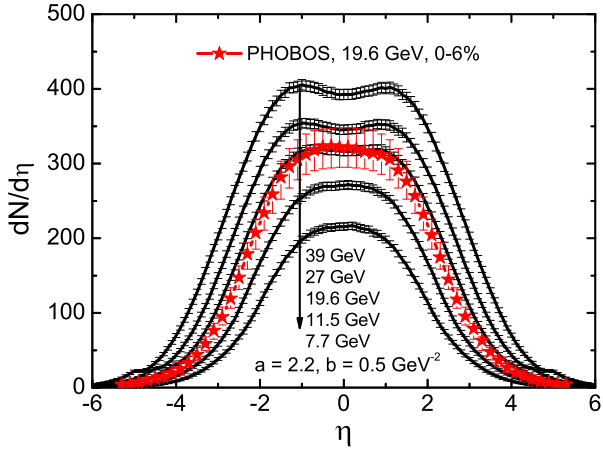}}
\caption{(Color online)
Pseudorapidity distributions of charged particles in central Au $+$ Au collisions at $\sqrt {s_{NN}} =$ 39, 27, 19.6, 11.5, and 7.7 GeV from the string melting AMPT model, compared with the PHOBOS data for Au $+$ Au collisions at $\sqrt {s_{NN}} =$ 19.6 GeV and centrality of 0-6$\%$ taken from Refs.~\cite{BA09,BA11}.}\label{etadis}
\end{figure}

We first reproduce the pseudorapidity distribution of charged particles by choosing suitable Lund string fragmentation parameters. As shown in Fig.~\ref{etadis}, we found the parameters $a = 2.2$ and  $b = 0.5$ GeV$^{-2}$ reproduce reasonably well the pseudorapidity distribution within $|\eta|<5$ in central Au+Au collisions at $\sqrt {s_{NN}}=19.6$ GeV measured by PHOBOS Collaboration. On the other hand, the pseudorapidity distribution is found to be insensitive to the parton scattering cross section. For the pseudorapidity distributions at other collision energies, there is no available experimental data so far.

\begin{figure}[htbp]
\centerline{\includegraphics[scale=0.3]{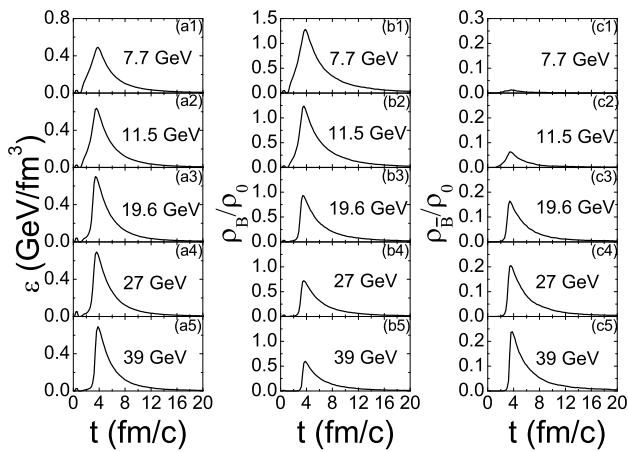}}
\caption{ Time evolution of the energy density (left column), the reduced baryon density (middle column), and the reduced antibaryon density (right column) in the central region of the hadronic phase in mid-central (20$-$30\%) Au + Au collisions at $\sqrt {s_{NN}} =$ 7.7, 11.5, 19.6, 27, and 39 GeV.}\label{density3}
\end{figure}

\begin{figure}[htbp]
\centerline{\includegraphics[scale=0.3]{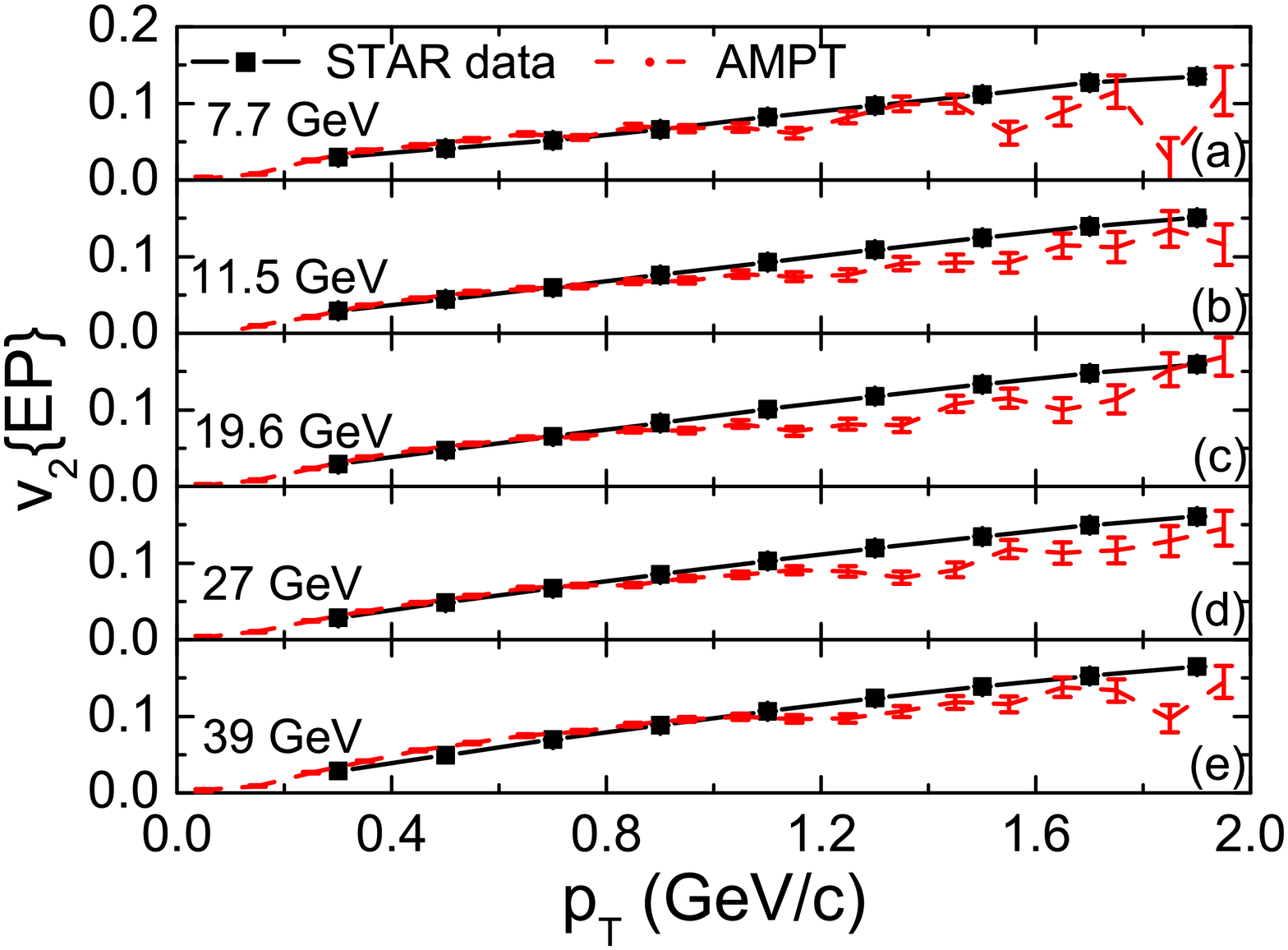}}
\caption{(Color online)
Transverse momentum ($p_{T}$) dependence of charged hadron elliptic flow ($v_{2}$)
at midpseudorapidities ($|\eta| < $ 1.0) in mid-central (20$-$30\%) Au + Au collisions at $\sqrt {s_{NN}} =$ 7.7 (a),
11.5 (b), 19.6 (c), 27 (d), and 39 GeV (e), compared with the STAR data taken from Ref.~\cite{L12STAR}.}
\label{v2EPrhicbes}
\end{figure}

The peak values of the energy density in the hadronic phase and the final elliptic flows of charged particles from the string-melting AMPT are fitted by the life time of the partonic phase and the parton scattering cross section. Using the baryon chemical potential and the temperature extracted from the statistical model~\cite{And10}, the energy densities at chemical freeze-out at $\sqrt {s_{NN}}=$ 7.7, 11.5, 19.6, 27, and 39 GeV are 0.49, 0.62, 0.68, 0.69, and 0.69 GeV/fm$^{3}$, from the hadron resonance gas model with the particle species evolved in ART/AMPT. By stopping the partonic phase at 2.75, 2.4, 2.15, 2.25, and 2.45 fm/c, respectively, we reproduce the peak values of the energy density in the central region of the hadronic phase as those from the hadron resonance gas model, as seen from the left column of Fig.~\ref{density3}. It is also seen in Fig.~\ref{density3} that the baryon density decreases with increasing collision energy, while the antibaryon density increases with increasing collision energy. Based on our model both the baryon and antibaryon densities are mostly lower than the saturation density $\rho_0=0.16$ fm$^{-3}$. In order to reproduce the charged particle elliptic flow at various RHIC-BES energies, we use the isotropic parton scattering cross sections of 3 mb for 7.7 GeV, 3 mb for 11.5 GeV, 6 mb for 19.6 GeV, 6 mb for 27 GeV, and 10 mb for 39 GeV. With the same event-plane method as applied in the experimental analysis~\cite{L12STAR,AMP98}, we reproduce reasonably well the charged hadron elliptic flow at midpseudorapidities in midcentral Au+Au collisions, as shown in Fig.~\ref{v2EPrhicbes}. We note that the HBT correlation is sensitive to the parton scattering cross section~\cite{ZWL02}, or equivalently, the shear viscosity of the partonic phase~\cite{AMPT1}. Also, in reality partons may also be affected by their mean-field potentials~\cite{JX14,Xu16}. The well fitted parton life time and the parton scattering cross section compensate other effects, since the energy density at chemical freeze-out and the experimental elliptic flow results need to be reproduced anyway.

\subsection{Pion femotoscopy and bulk properties of emission source}

\begin{figure}[htbp]
\centerline{\includegraphics[scale=0.32]{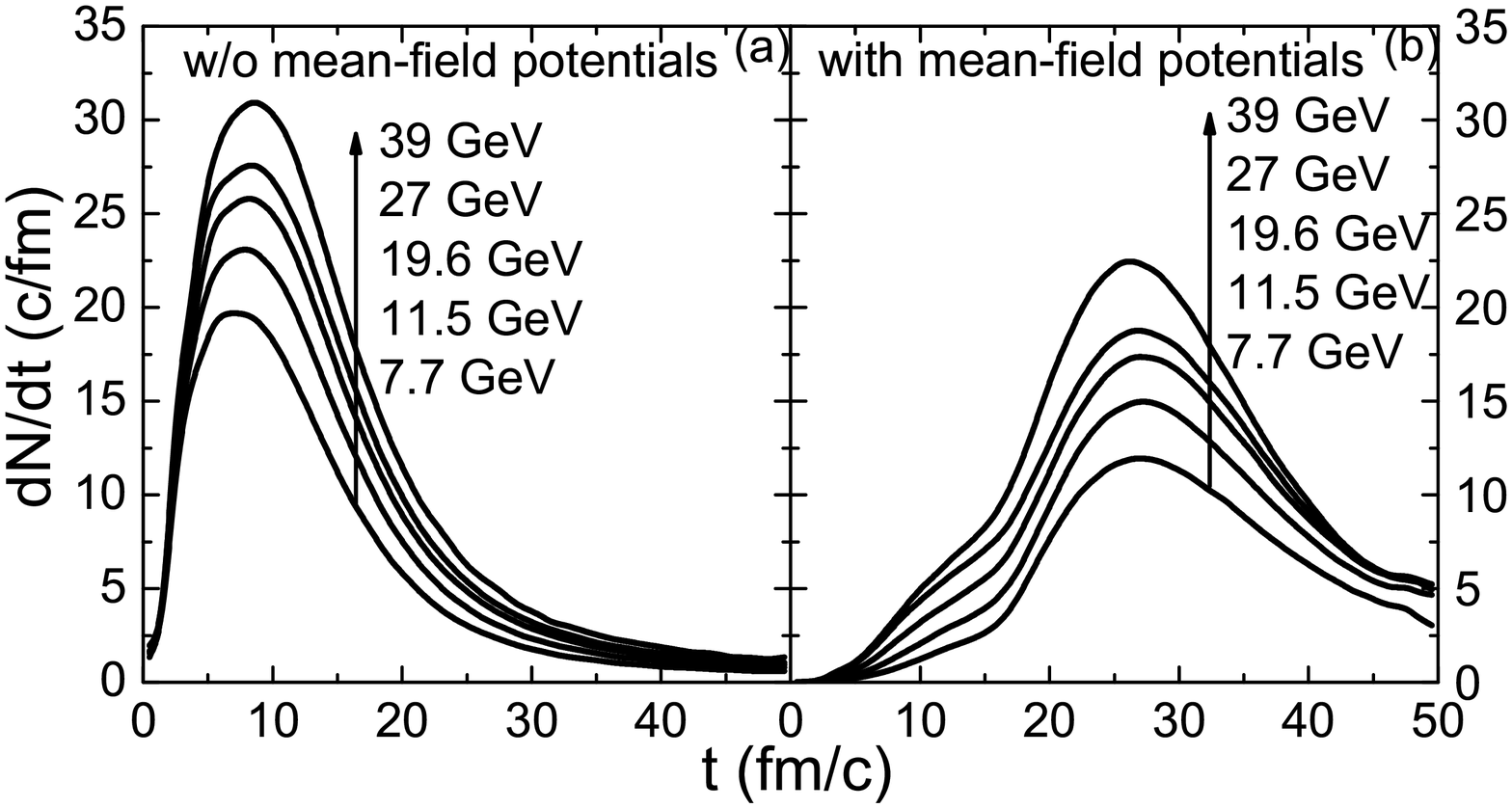}}
\caption{ Distributions of hadron freeze-out time with (right) or without (left) mean-field potentials in Au + Au
collisions at $\sqrt {s_{NN}} =$ 7.7, 11.5, 19.6, 27, and 39 GeV.}\label{dndt4}
\end{figure}

With the string-melting AMPT calibrated above, we first investigate
the effects of the hadronic mean-field potentials on the bulk
properties of the emission source. Using the $``$free-streaming$"$
criterion for the kinetic freeze-out of hadrons as mentioned in
Sec.~\ref{ampt}, we found that freeze-out times for hadrons are
generally much later with mean-field potentials compared to the case
without mean-field potentials, as shown in Fig.~\ref{dndt4}. This is
understandable since the $``$soft$"$ attractive mean-field
potentials at lower densities, especially for pions and baryons,
delay the emission of these particles, so the system freezes out
kinetically at a much later time on average. In addition, the
mean-field potentials lead to a broader windows of the emission time
and presumedly weaker correlations.

\begin{figure}[htbp]
\centerline{\includegraphics[scale=0.3]{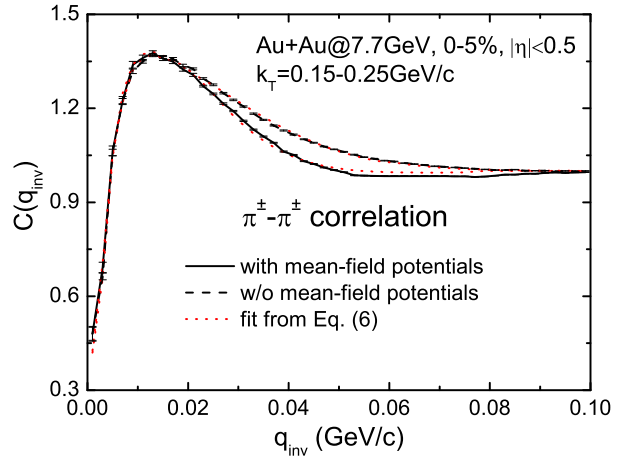}}
\caption{(Color online) HBT correlation of charged pions at midpseudorapidities with and without mean-field potentials in central Au + Au collisions at $\sqrt {s_{NN}} =$ 7.7 GeV.}\label{1dfit6}
\end{figure}

\begin{table}

%  \centering
  \fontsize{6.5}{8}\selectfont
%  \begin{threeparttable}
  \caption{Comparison of HBT parameters obtained by the three-dimensional fit of Eq.~(\ref{3d}) without mean-field potentials, with mean-field potentials, and those from STAR data analyses~\cite{LA15} at various collision energies. }
  \label{tab:performance_comparison}
    \begin{tabular}{|c|c|c|c|c|c|}
    \hline
    \multicolumn{2}{c|}{$\sqrt{s^{}_{NN}}$ (GeV)}
    &$\lambda$&$R_{o}$ (fm)&$R_{s}$ (fm)&$R_{l}$ (fm) \cr\hline
    \multirow{3}{*}{7.7}
    &cascade&    $0.673 \pm 0.007$& $5.58 \pm 0.03$&    $4.75 \pm 0.03$&     $4.39 \pm 0.03$ \cr
    &mean-field& $0.719\pm0.007$ &   $6.16 \pm 0.04$ &   $5.53 \pm 0.04$ &    $5.51 \pm 0.04$ \cr
    &expt.&      $0.532 \pm 0.007$&   $5.57 \pm 0.13$&      $4.93 \pm 0.10$&    $5.01 \pm 0.11$ \cr\hline

     \multirow{3}{*}{11.5}
     &cascade&   $0.663\pm0.007$&   $5.59 \pm 0.03$&   $4.72 \pm 0.03$&    $4.39 \pm 0.03$ \cr
     &mean-field&$0.704\pm0.007$&    $6.33 \pm 0.04$&   $5.54 \pm 0.04$&    $5.67 \pm 0.04$ \cr
     &expt.&     $0.508 \pm 0.004$& $5.68 \pm 0.07$&    $4.79 \pm 0.05$&    $5.43 \pm 0.07$ \cr\hline

     \multirow{3}{*}{19.6}
     &cascade&    $0.656\pm0.007$ &  $5.60 \pm 0.03$&   $4.78 \pm0.03$&      $4.43 \pm 0.03$ \cr
     &mean-field&$0.698\pm0.008$&     $ 6.39 \pm 0.04$&   $5.48 \pm 0.04$&    $5.76 \pm 0.04$ \cr
     &expt.&     $0.498 \pm 0.002$&  $5.84 \pm 0.05$&   $4.84 \pm 0.03$&         $5.80  \pm 0.05$ \cr\hline

     \multirow{3}{*}{27}
     &cascade&   $0.651 \pm 0.007$&    $5.60 \pm 0.03$&   $4.78 \pm 0.03$ &   $4.49 \pm 0.03$ \cr
     &mean-field&$0.689 \pm0.008$ &    $6.41 \pm 0.04$&   $5.51 \pm 0.04$ &   $5.88 \pm 0.04$ \cr
     &expt.&     $0.492 \pm 0.002$&    $5.82 \pm 0.03$&   $4.89  \pm 0.02$ &   $5.99 \pm0.04$ \cr\hline

     \multirow{3}{*}{39}
     &cascade&    $0.655 \pm 0.007$&   $5.63  \pm 0.03$&   $4.79\pm 0.03$&   $4.55 \pm 0.03$\cr
     &mean-field& $0.678 \pm 0.008$&   $6.42 \pm 0.04$&   $5.53 \pm 0.04$&   $5.95 \pm 0.04$\cr
     &expt.&      $0.491 \pm 0.004$&   $5.86   \pm 0.07$&   $4.97 \pm 0.05$&   $6.18 \pm 0.08$\cr\hline
    %\hline

    %\hline

    \end{tabular}
%    \end{threeparttable}
\end{table}

Pion interferometry is a useful tool to investigate bulk properties
of the hot and dense matter formed in relativistic heavy-ion
collisions. The HBT correlations for charged pions
($\pi^{\pm}$-$\pi^{\pm}$) with and without mean-field potentials are
compared in Fig.~\ref{1dfit6}. It is seen that the correlation
functions have the same peak while it is sharper with mean-field
potentials, since the correlation at larger $q_{\rm inv}$ is
suppressed, as a result of later and longer duration of the
emission. From the fit by
Eq.~(\ref{1d}), the direction-averaged radius of the
emission source $R_{\rm inv}$ is expected to be larger with
mean-field potentials due to the shaper correlation function
compared to that without mean-field potentials. In the former case,
the larger radius of the emission source is also consistent with the
later emission as shown in Fig.~\ref{dndt4}. We note that the difference between the $\pi^+-\pi^+$ correlation and the $\pi^--\pi^-$ correlation is found to be small due to the small isospin asymmetry in the hadronic phase.

In the three-dimensional HBT analysis, we apply the $\chi^2$ fit to
Eq.~(\ref{3d}) and obtain the chaoticity parameter
$\lambda$ and the outward, sideward, and longitudinal HBT radii,
i.e., $R_{o}$, $R_{s}$, and $R_{l}$. Table. I compares the resulting
HBT parameters with and without mean-field potentials and those from
STAR data analyses~\cite{LA15}. It is found that the values of
$\lambda$ from AMPT are generally larger than the STAR results, with
or without mean-field potentials. With mean-field potentials, the
values of $R_l$ are more consistent with STAR results, while the
values of $R_o$ and $R_s$ are overestimated. As is well
known~\cite{UH99,FR04,MAL05}, a stronger transverse flow reduces the
homogeneity lengths and thus the HBT radii. Since the AMPT model
underestimates the transverse flow due to the massless partons,
incorporating the hadronic mean-field potentials and enhancing the
transverse flow are promising to reproduce better the radii of the
emission source extracted from experimental measurements.

\begin{figure}[htbp]
\centerline{\includegraphics[scale=0.32]{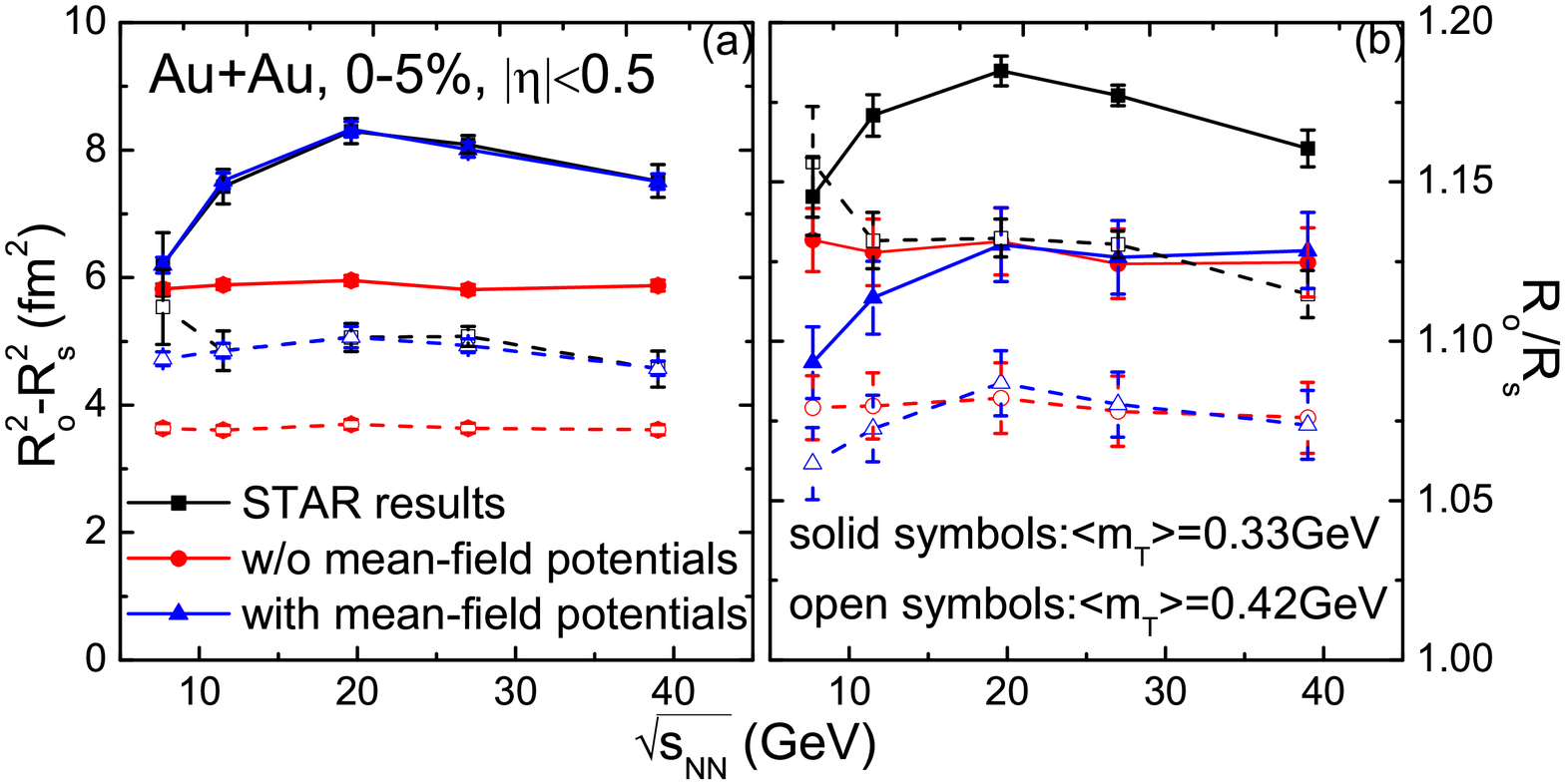}}
\caption{(Color online) Collision energy dependence of $R^2_{o}-R^2_{s}$ and $R_{o}/ R_{s}$ extracted from freeze-out charged pions with and without mean-field potentials in central Au+Au collisions compared with those extracted from STAR measurements~\cite{LA15}.}\label{outside}
\end{figure}

The quantity $R^{2}_{o} - R^{2}_{s}$ is of special interest as it is related to the emission duration $\Delta t$ in the limit of a static source through the empirical relation
$R^{2}_{o} - R^{2}_{s} \approx \beta_{T}^{2} \Delta t^{2}$, where $\beta_{T}=k_T/m_T$ is the particle speed in the source rest frame with $m_T=\sqrt{k_T^2+m_\pi^2}$. It was argued~\cite{Lac15} that the finite-size scaling analysis on the peak of $R^{2}_{o}$ - $R^{2}_{s}$ as a function of the C.M. energy $\sqrt {s_{NN}}$ at various centralities may reveal the critical point in the QCD phase diagram. The ratio $R_{o}/ R_{s}$ was first proposed in Ref.~\cite{DH96}, and it has the advantage of removing the overall scale of the system. Based on the hydrodynamics calculation~\cite{DH96}, $R_{o}/ R_{s}$ exhibits a peak near the onset of deconfinement. Figure~\ref{outside} compares the $\sqrt {s_{NN}}$ dependence of $R^{2}_{o}$ - $R^{2}_{s}$ and $R_{o}/ R_{s}$ with or without mean-field potentials with that from the STAR results. It is seen that the peaks around $\sqrt {s_{NN}} \approx $ 20 GeV for both $R^{2}_{o}$ - $R^{2}_{s}$ and $R_{o}/ R_{s}$ are reproduced with mean-field potentials, but there is no such peak without mean-field potentials. Furthermore, results of $R^{2}_{o}$ - $R^{2}_{s}$ with mean-field potentials reproduce better the STAR data while results of $R_{o}/ R_{s}$ are generally underestimated from AMPT calculations compared with the STAR data.

\begin{figure}[htbp]
\centerline{\includegraphics[scale=0.3]{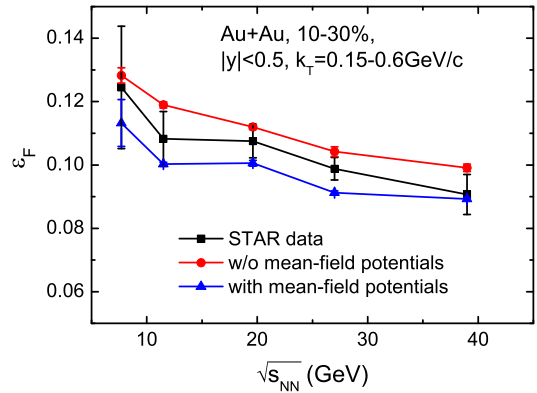}}
\caption{(Color online) Freeze-out eccentricity $\epsilon_F^{}$ extracted from the HBT correlation of charged pions with and without mean-field potentials as a function of the C.M. energy $\sqrt{s_{NN}}$ in midcentral Au + Au collisions compared with those extracted from STAR measurements~\cite{LA15}. }\label{eccentricity8}
\end{figure}

From both hydrodynamics~\cite{Kol03} and transport
model~\cite{Lis11} studies, the eccentricity of the emission source
is related to the equation of state and the order of quark-hadron
phase transition, and it can be obtained from the HBT analysis
through the relation $\epsilon_{F}^{}  \approx  2
{R_{s,2}^{2}}/{R_{s,0}^{2}}$~\cite{FR04}, where $R_{s,0}$ and
$R_{s,2}$ are respectively the zeroth- and second-order coefficients
from fitting $R_s$ at azimuthal angles $\Phi=\pi/4$, $\pi/2$,
$3\pi/4$, and $\pi$ according to Eq.~(\ref{az}) up to the second
order. Figure~\ref{eccentricity8} compares the C.M. energy $\sqrt
{s_{NN}}$ dependence of $\epsilon_{F}^{}$ with and without
mean-field potentials with that from STAR analyses. The
$\epsilon_{F}^{}$ values at various collision energies are smaller
with mean-field potentials as a result of later freeze-out of the
system as shown in Fig.~\ref{dndt4}, and the STAR results are
in-between the $\epsilon_{F}^{}$ values with and without mean-field
potentials. The seemingly constant $\epsilon_F^{}$ from
$\sqrt{s_{NN}}=11.5$ to 19.6 GeV from the STAR result is also
observed in the scenario with mean-field potentials.

\subsection{HBT correlations of protons, kaons, and antiprotons}

\begin{figure}[htbp]
\centerline{\includegraphics[scale=0.32]{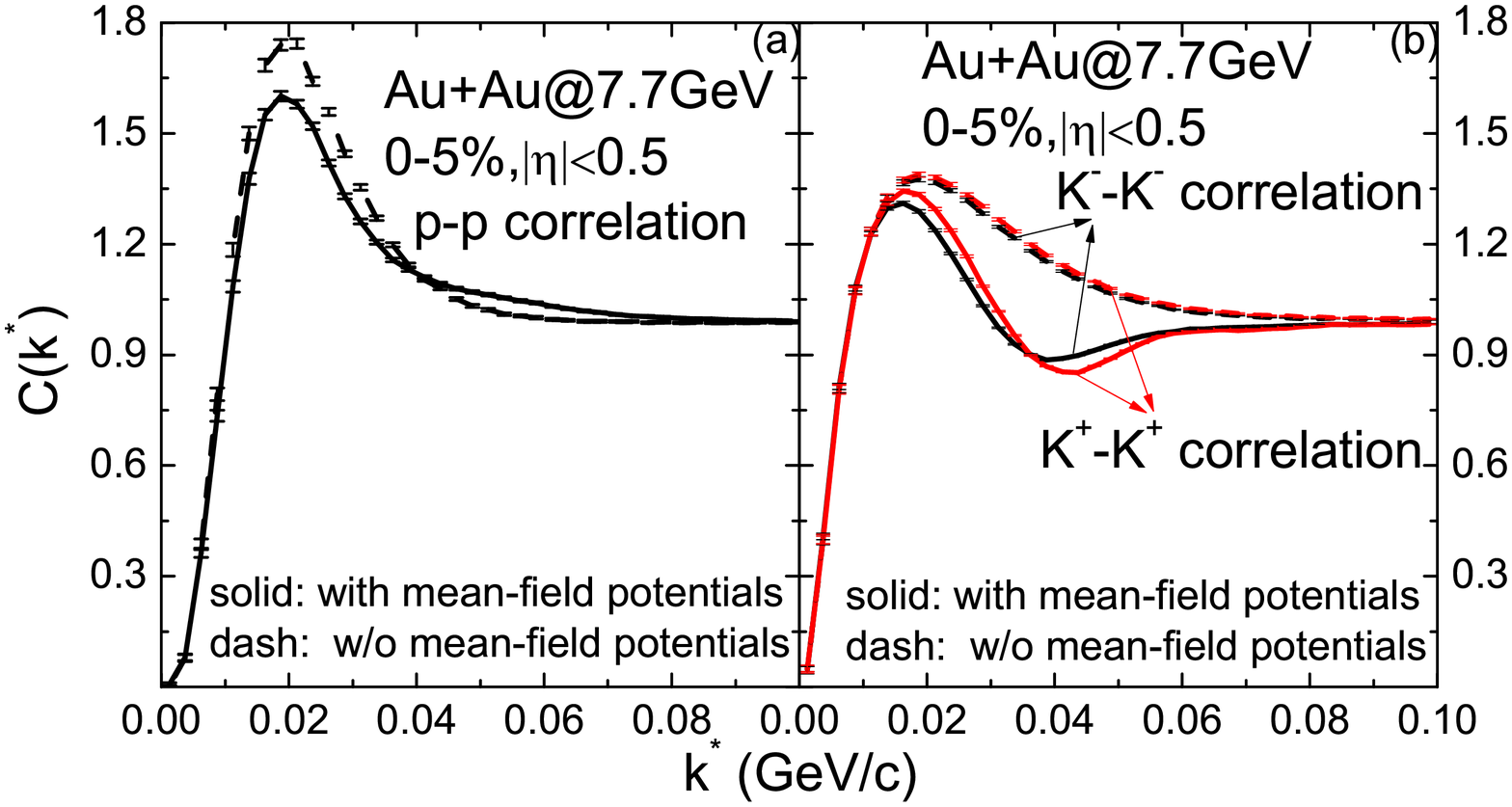}}
\caption{(Color online) Proton-proton HBT correlation (left) and charged kaon HBT correlation (right) at midpseudorapidities with and without mean-field potentials in central Au+Au collisions at $\sqrt {s_{NN}} =$ 7.7 GeV. } \label{protonkaon5}
\end{figure}

The HBT correlation is not only useful for extracting bulk properties of the emission source but valuable in obtaining the information of the mean-field potentials for particles such as protons or kaons as well as their antiparticles. The left panel of Fig.~\ref{protonkaon5} displays the proton-proton correlation with and without mean-field potentials in central Au+Au collisions at $\sqrt{s_{NN}}=7.7$ GeV. The peak at about $k^*= 0.02$ GeV/c is due to the strong final-state $s$-wave attraction, while the correlation is suppressed at
smaller $k^*$ as a result of the Coulomb repulsion between two protons and the anti-symmetrized wave function of the identical particle pair. With mean-field potentials, the proton-proton correlation function is suppressed at smaller $k^*$ but enhanced at larger $k^*$, indicating a suppression of energetic emissions but an enhancement of thermal emissions. From fitting the proton-proton HBT correlation according to Eq.~(\ref{1d}), we found that the direction-averaged emission radius for proton is also larger with mean-field potentials compared to that without mean-field potentials. Kaons are not much affected by resonance decays, so their correlation function has advantages over pions in some aspects~\cite{Sof02,PHENIX09}. As shown in the right panel of Fig.~\ref{protonkaon5}, kaon interferometry is affected by their mean-field potentials. Typically, the correlation is much suppressed at intermediate $k^*$ with kaon potentials due to the later and longer duration of the emission, and the splitting of the $K^+-K^+$ and $K^--K^-$ correlation is clearly observed due to their different mean-field potentials in baryon-rich hadronic matter. We have further found that the kaon correlation function is sensitive to kaon potentials but rather insensitive to the potentials for protons or pions.

\begin{figure}[htbp]
\centerline{\includegraphics[scale=0.3]{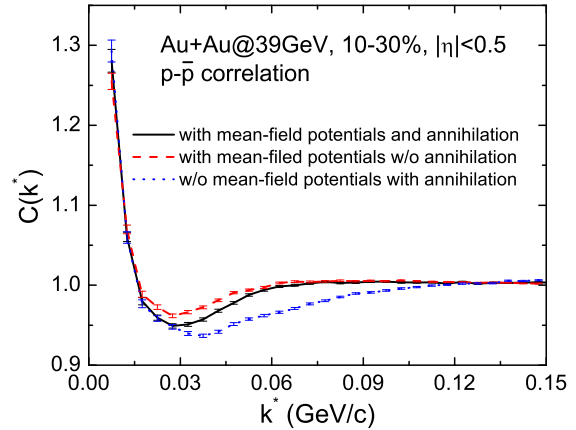}}
\caption{(Color online) Proton-antiproton HBT correlation with and/or without mean-field potentials and baryon-antibaryon annihilations in midcentral Au+Au collision at $\sqrt{s_{NN}}=39$ GeV.} \label{annihilation9}
\end{figure}

In the afterburner part after kinetic freeze-out, the baryon-antibaryon annihilation has been taken into account in the HBT analysis~\cite{LL82}, and used in the experimental studies of baryon-antibaryon correlations~\cite{MPS13,MPS14}. In the hadronic evolution described by the ART/AMPT model, the baryon-antibaryon annihilation cross sections are related to their branching ratios to different multipion states~\cite{KO87,Lin05}, and the inverse processes are incorporated by the detailed balance condition.  It will be interesting to investigate the proton-antiproton correlation from the interplay of their mean-field potentials and annihilation effect in transport simulations. In order to illustrate the two effects independently, we display in Fig.~\ref{annihilation9} the event-by-event proton-antiproton correlation by turning on and off their mean-field potentials as well as the annihilation and inverse process separately. The strong correlation at lower $k^*$ ($k^*<0.015$ GeV/c) is due to the attractive Coulomb interaction. By comparing the correlation in different scenarios, although it is still difficult to reproduce the experimental results quantitatively, we can draw a qualitative conclusion that the attractive potential between protons and antiprotons as well as the larger emission source due to mean-field potentials leads to a weaker anti-correlation, while the annihilation leads to a stronger anti-correlation.

\section{Conclusions}
\label{summary}

In the present work, we have studied effects of hadronic mean-field potentials, which always exist in the hadronic phase of relativistic heavy-ion collisions, on the HBT correlation based on the framework of a multiphase transport model, with parameters calibrated in order to reproduce the particle multiplicity, the energy density at chemical freeze-out, and the charged particle elliptic flow at RHIC-BES energies. Generally, the hadronic mean-field potentials delay the kinetic freeze-out of the system and enlarge the HBT radii, and they may affect the collision energy dependence of $R_o^2-R_s^2$ and $R_o/R_s$ as well as the eccentricity of the emission source extracted from pion interferometric analyses. On the other hand, the HBT correlations for protons, kaons, and antiprotons can be useful in extracting information of their mean-field potentials in the baryon-rich hadronic matter as well as understanding baryon-antibaryon annihilations. Our study is useful in understanding the dynamics in relativistic heavy-ion collisions at RHIC-BES energies relevant for mapping out the QCD phase diagram.

\begin{acknowledgments}
We thank Chen Zhong for maintaining the high-quality performance of the computer facility, and acknowledge helpful discussions with Zheng-Qiao Zhang and communications with R. Lednick$\acute{y}$ and $\L$ukasz Kamil Graczykowski. This work was supported by the Major State Basic Research Development Program (973 Program) of China under Contract Nos. 2015CB856904 and 2014CB845401, the National Natural Science Foundation of China under Grant Nos. 11475243 and 11421505, the "100-talent plan" of Shanghai Institute of Applied Physics under Grant Nos. Y290061011 and Y526011011 from the Chinese Academy of Sciences, and the Shanghai Key Laboratory of Particle Physics and Cosmology under Grant No. 15DZ2272100.
\end{acknowledgments}

\appendices
\renewcommand\thesection{APPENDIX~\Alph{section}}
\renewcommand\theequation{\Alph{section}.\arabic{equation}}
\section{femotoscopy theory}
\label{LLAM}

In this appendix, we briefly remind the reader of the main formulaes
given by Lednicky and Lyuboshitz~\cite{LL82,RL06,RL08,RL09}, which
are used in the calculation of the HBT correlation and relevant for
the understanding of physics in the present work. In such framework,
the particle correlations at small relative velocities are sensitive
to the space$-$time characteristics of the production processes on a
femtometer scale owing to the effects of quantum statistics and
final-state interactions. Effects of the Coulomb interaction are
expected to dominate the correlations of charged particles at the
relative momenta in the two-particle rest frame smaller than the
inverse Bohr radius of the two-particle system, suppressing or
enhancing the production of particles with like or unlike charges,
respectively. The correlation function is then given by a square of
the properly symmetrized Bethe-Salpeter amplitude representing the
continuous spectrum of the two-particle states, averaging over the
four-coordinates of the emitters and the total spin of the
two-particle system. In the femotoscopy theory, an assumption is
made that the FSI of the particle pairs is independent of their
production.

The two-particle correlation function can be written as
\begin{eqnarray}
C(\textbf{k}^{*}) = \frac{\int S(\textbf{r}^{*},\textbf{k}^{*})\left |\Psi _{\textbf{k}^{*}}( \textbf{r}^{*})  \right |^{2}d^4\textbf{r}^{*}}{\int S(\textbf{r}^{*},\textbf{k}^{*})d^4\textbf{r}^{*}}. \label{fc}
\label{A1}
\end{eqnarray}%
In the above, $\textbf{r}^{*} = \textbf{x}_{1} - \textbf{x}_{2}$ is a relative space-time separation of the two particles at their kinetic freeze-out, $\textbf{k}^{*}$ is the momentum of the first particle in the pair rest frame, $S(\textbf{r}^{*},\textbf{k}^{*})$ is the source emission function interpreted as the probability to emit a particle pair with given $\textbf{r}^{*}$ and $\textbf{k}^{*}$, and $\Psi _{\textbf{k}^{*}}( \textbf{r}^{*})$ is the Bethe-Salpeter amplitude.

In the region of interest with small $\textbf{k}^{*}$, the short-range particle interaction that introduces correlation is dominated by the $s$-wave interaction, whose interaction range is usually small compared with the distance $\textbf{r}^{*}$ between the particle pair in their C.M. system but large compared with the strong interaction range. In this limit, the asymptotic solution of the wave function of the two charged particles can be approximately written as
%\begin{footnotesize}
\begin{eqnarray}
\Psi_{\textbf{k}^{*}}( \textbf{r}^{*}) &=& e^{i\delta_{c}}\sqrt{A_{c}(\eta)} \notag \\
&\times&\left[ e^{-i\textbf{k}^{*}\textbf{r}^{*}}F(-i\eta,1,i\xi)
\!+\! f_{c}(\textbf{k}^{*})\frac{\tilde{G}(\rho ,\eta)}{\textbf{r}^{*}}\right]. \notag
\end{eqnarray}%
%\end{footnotesize}
In the above, $A_{c}(\eta) = 2\pi\eta\left [ \exp\left ( 2\pi\eta \right )-1 \right
]^{-1}$ is the Coulomb penetration factor, with $\eta = (k^{*}a_{c})^{-1}$ where $a_{c}$ is the two-particle
Bohr radius including the sign of the interaction, $\delta_{c} = arg\Gamma (1+i\eta)$ is the Coulomb $s$-wave phase shift, $F(-i\eta,1,i\xi) = 1+(-i\eta) (i\xi)/1!^{2} + (-i\eta) (-i\eta +1)(i\xi)^{2}/2!^{2}+...$ is the confluent hypergeometric function with $\xi =\textbf{k}^{*}\textbf{r}^{*} +k^{*}r^{*}$, and
$\tilde{G}(\rho ,\eta) = \sqrt{A_{c}}[G_{0}(\rho ,\eta)+iF_{0}(\rho ,\eta)]$ is a combination of regular [$F_{0}(\rho ,\eta)$] and singular [$G_{0}(\rho ,\eta)$] $s$-wave Coulomb functions, whose detailed forms can be found in Refs.~\cite{RL09, MG86}, with $\rho=k^{*}r^{*}$.
%\begin{footnotesize}
%\begin{eqnarray}
$f_{c}(k^{*})=\left [ \frac{1}{f_{0}}+\frac{1}{2}d_{0}{k^*}^{2}-\frac{2}{a_{c}}h(\eta)-ik^{*}A_{c}(\eta) \right ]^{-1}$
%\end{eqnarray}%
%\end{footnotesize}
is the amplitude of the $s$-wave elastic scattering due to the short-range interaction renormalized by the long-range Coulomb
forces, with $h(\eta)=\eta^{2}\sum_{n=1}^{\infty}\left [ n(n^{2}+\eta^{2})\right ]^{-1}- C - \ln\left | \eta \right |$
where $C \doteq 0.5772$ is the Euler constant, $f_{0}$ being the scattering length, and $d_{0}$ being the effective radius of
the strong interaction. Both $f_{0}$ and $d_{0}$ are essential parameters characterizing the main properties of the final-state strong interaction, and can be extracted from the correlation function measured experimentally~\cite{LA15N}. The imaginary part of $f_{0}$ corresponds to the annihilation
process~\cite{MPS14,JS13}, while the interaction cross section can be expressed in terms of the scattering amplitude as $\sigma =4\pi\left | f_{c}(k^{*}) \right |^{2}$~\cite{RL09,AKH14} in the effective range approximation mentioned above.


\begin{thebibliography}{99}

%BES
\bibitem{MMA10}  M. M. Aggarwal {\it et al.} (STAR Collaboration), arXiv: 1007.2613 [nucl-ex].
\bibitem{JX12}   J. Xu, L. W. Chen, C. M. Ko, and Z. W. Lin, Phys. Rev. C \textbf{85}, 041901(R) (2012).
%HBT orginals
\bibitem{HBT56}  R. H. Brown and R. Q. Twiss, Nature \textbf{178}, 1056 (1956).
\bibitem{MSH86}  M. S. Hybertsen and S. G. Louie, Phys. Rev. B \textbf{34}, 8 (1986).
\bibitem{MS99}   M. Henny, S. Oberholzer, C. Strunk, T. Henzel, K. Ensslin, M. Holland, and
C. Schonenberger, Science \textbf{284}, 296 (1999).
\bibitem{SM05}   M. Schellekens {\it et al.}, Science \textbf{310}, 648 (2005).
%HBT HIC reviews
\bibitem{Boa90} D. H. Boal, C. K. Gelbke, and B. K. Jennings, Rev. Mod. Phys. \textbf{62}, 553 (1990).
\bibitem{WB92}   W. Bauer, C. K. Gelbke, and S. Pratt, Annu. Rev. Nucl. Part. Sci. \textbf{42}, 77 (1992).
\bibitem{UAW99}  U. A. Wiedemann and U. Heinz, Phys. Rep. \textbf{319}, 145 (1999).
\bibitem{UH99}   U. Heinz and B. V. Jacak, Ann. Rev. Nucl. Part. Sci. \textbf{49}, 529 (1999).
\bibitem{Wei00} R. M. Weiner, Phys. Rep. \textbf{327}, 249 (2000).
\bibitem{MAL05} M. A. Lisa, S. Pratt, R. Soltz, and U. Wiedemann, Ann. Rev. Nucl. Part. Sci. \textbf{55}, 357 (2005).
%HBT and QGP evolution
\bibitem{SP84}   S. Pratt, Phys. Rev. Lett. \textbf{53}, 1219 (1984).
\bibitem{SP86}   S. Pratt, Phys. Rev. D \textbf{33}, 1314 (1986).
\bibitem{GB88}   G. Bertsch {\it et al.}, Phys. Rev. C \textbf{37}, 1896 (1988).
\bibitem{SP90a}   S. Pratt {\it et al.}, Phys. Rev. C \textbf{42}, 2646 (1990).
\bibitem{DH96}   D. H. Rischke and M. Gyulassy, Nucl. Phys. A \textbf{698}, 603 (1996).
%mean-field potential effect on HBT
\bibitem{LQF08} Q. F. Li, M. Bleicher, and H. St\"ocker, Phys. Lett. B \textbf{659}, 525 (2008).
\bibitem{CLW03} L. W. Chen, V. Greco, C. M. Ko, and B. A. Li, Phys. Rev. Lett. \textbf{90}, 162701 (2003).
%ampt
\bibitem{Lin05}  Z. W. Lin, C. M. Ko, B. A. Li, B. Zhang, and S. Pal, Phys. Rev. C \textbf{72}, 064901 (2005).
\bibitem{Wang91}  X. N. Wang and M. Gyulassy, Phys. Rev. D \textbf{44}, 3501 (1991).
\bibitem{Zhang98}  B. Zhang, Comp. Phys. Comm. \textbf{109}, 193 (1998).
\bibitem{Li95}   B. A. Li and C. M. Ko, Phys. Rev. C \textbf{52}, 2037 (1995).
\bibitem{Xu16} J. Xu and C. M. Ko, Phys. Rev. C \textbf{94}, 054906 (2016).
%test-particle method
\bibitem{Won82} C. Y. Wong, Phys. Rev. C \textbf{25}, 1460 (1982).
%hadronic mean-field potentials
\bibitem{GQ94}   G. Q. Li, C. M. Ko, X. S. Fang, and Y. M. Zheng, Phys. Rev. C \textbf{49}, 1139 (1994).
\bibitem{GQ97}   G. Q. Li, C. H. Lee, and G. E. Brown, Nucl. Phys. A \textbf{625}, 372 (1997).
\bibitem{NK01}   N. Kaiser and W. Weise, Phys. Lett. B \textbf{512}, 283 (2001).
\bibitem{GE75}   G. E. Brown and W. Weise, Phys. Rep. \textbf{22}, 279 (1975).
\bibitem{JX81}   J. Xu, C. M. Ko, and Y. Oh, Phys. Rev. C \textbf{81}, 024910 (2010); J. Xu {\it et al.}, Phys. Rev. C \textbf{87}, 067601 (2013).
\bibitem{LX93}   L. Xiong, C. M. Ko, and V. Koch, Phys. Rev. C \textbf{47}, 788 (1993).
\bibitem{Zha17} Z. Zhang and C. M. Ko, Phys. Rev. C \textbf{95}, 064604 (2017).

\bibitem{CRAB}   S. Pratt {\it et al.}, Nucl. Phys. A \textbf{566}, 103c (1994).
\bibitem{LL82}   R. Lednick$\acute{y}$ and V. L. Lyuboshitz, Sov. J. Nucl. Phys. \textbf{35}, 770 (1982).
\bibitem{JA05}   J. Adams {\it et al.} (STAR Collaboration), Phys. Rev. C \textbf{71}, 044906 (2005).
\bibitem{YMS98}  Y. M. Sinyukov, R. Lednicky, S. V. Akkelin, J. Pluta, and B. Erazmus, Phys. Lett. B \textbf{432}, 248 (1998).
\bibitem{MGB91}  M. G. Bowler, Phys. Lett. B \textbf{270}, 69 (1991).

\bibitem{BA09}   B. Alver {\it et al.} (PHOBOS Collaboration), Phys. Rev. Lett. \textbf{102}, 142301 (2009).
\bibitem{BA11}   B. Alver {\it et al.} (PHOBOS Collaboration), Phys. Rev. Lett. \textbf{82}, 024913 (2011).

%statistical model
\bibitem{And10} A. Andronica, P. Braun-Munzingera, and J. Stachel,
Nucl. Phys. A \textbf{834}, 237c (2010).

\bibitem{L12STAR} L. Adamczyk {\it et al.} (STAR Collaboration), Phys. Rev. C \textbf{86}, 054908 (2012).
\bibitem{AMP98}  A. M. Poskanzer and S. A. Voloshin, Phys. Rev. C \textbf{58}, 1671 (1998).

\bibitem{ZWL02}  Z. W. Lin, C. M. Ko, and S. Pal, Phys. Rev. Lett \textbf{89}, 152301 (2002).

\bibitem{AMPT1}  Y. Zhang, Y. B. Zhang, J. L. Liu, and L. Huo, Phys. Rev. C \textbf{92}, 014909 (2015).

\bibitem{JX14}   J. Xu, T. Song, C. M. Ko, and F. Li, Phys. Rev. Lett. \textbf{112}, 012301 (2014).

\bibitem{LA15}   L. Adamczyk {\it et al.} (STAR Collaboration), Phys. Rev. C \textbf{92}, 014904 (2015).

\bibitem{FR04}   F. Reti$\grave{e}$re and M. A. Lisa, Phys. Rev. C \textbf{70}, 044907 (2004).

%o-s
\bibitem{Lac15} R. A. Lacey, Phys. Rev. Lett. \textbf{114}, 142301 (2015).

%eccentricity
\bibitem{Kol03} P. F. Kolb and U. Heinz, arXiv: nucl-th/0305084.
\bibitem{Lis11} M. A. Lisa, E. Frodermann, G. Graef, M. Mitrovski, E. Mount,
H. Petersen, and M. Bleicher, New Journal of Physics \textbf{13}, 065006 (2011).
%kaon
\bibitem{Sof02} S. Soff, S. A. Bass, D. H. Hardtke, and S. Y. Panitkin, Phys. Rev. Lett. \textbf{88}, 072301 (2002).
\bibitem{PHENIX09} S. Afanasiev {\it et al.}, Phys. Rev. Lett. \textbf{103}, 142301 (2009)

%annihilation
%\bibitem{GM79}   M. Gyulassy, S. K. Kauffmann, and L. W. Wilson, Phys. Rev. C \textbf{20}, 2267 (1979).
\bibitem{MPS13}  M. P. Szyma$\acute{n}$ski, Nucl. Phys. A \textbf{904}, 447c (2013).
%\bibitem{TGL16}  T. G. Lee and C. Y. Wong,  Phys. Rev. C \textbf{93}, 014601 (2016).
%\bibitem{GG60}   G. Goldhaber, S. Goldhaber, W. Y. Lee, and A. Pais, Phys. Rev \textbf{120}, 301 (1960).
\bibitem{MPS14}  M. P. Szyma$\acute{n}$ski, arXiv: 1403.0462 [nucl-ex].
\bibitem{KO87}   C. M. Ko and R. Yuan, Phys. Lett. B \textbf{192}, 31 (1987).

\bibitem{RL06}   R. Lednick$\acute{y}$, Nucl. Phys. A \textbf{774}, 189 (2006).
\bibitem{RL09}   R. Lednick$\acute{y}$, Phys. of Part. and Nucl. \textbf{40}, 307 (2009).
\bibitem{RL08}   R. Lednick$\acute{y}$, Phys. of Ato. Nucl. \textbf{71}, 1572 (2008).
\bibitem{MG86} M. Gmitro, J. Kvasil, R. Lednick$\acute{y}$, and V. L. Lyuboshitz, Czech. J. Phys. B \textbf{36}, 1281 (1986).
\bibitem{LA15N}  L. Adamczyk {\it et al.} (STAR Collaboration), Nature \textbf{527}, 347 (2015).
\bibitem{JS13}   J. Salzwedel, arXiv: 1305.4962 [nucl-ex].

\bibitem{AKH14}  A. Kisiel, H. Zbroszczyk, and M. Szyma$\acute{n}$ski, Phys. Rev. C \textbf{89}, 054916 (2014).

\end{thebibliography}
\end{document}